\documentclass[journal]{IEEEtran}
\usepackage{graphicx}
\usepackage{amsmath}
\usepackage{cite}
\usepackage{booktabs}
\usepackage{multirow}
\usepackage{array}
\usepackage{tabularx}
\usepackage{afterpage}
\usepackage[export]{adjustbox}
\usepackage{xcolor}
\usepackage{siunitx}

\title{Automated ECG Interval Measurement and Wave Delineation Using\\
Fast Fourier Convolution ResNet}

\author{Farhan~Adam~Mukadam\IEEEauthorrefmark{1,*}\IEEEauthorrefmark{5},
Harshit~Mishra\IEEEauthorrefmark{2},
Nachiket~Makwana\IEEEauthorrefmark{2},
Pradyot~Tiwari\IEEEauthorrefmark{3},
Subramani~Kandasamy\IEEEauthorrefmark{4},
K.\,V.\,S.~Hari\IEEEauthorrefmark{1}\\[1ex]
\IEEEauthorrefmark{1}Indian Institute of Science, Bengaluru, India\\
\IEEEauthorrefmark{2}Gauze, India\\
\IEEEauthorrefmark{3}Waikato District Health Board, Hamilton, New Zealand\\
\IEEEauthorrefmark{4}Christian Medical College, Vellore, India\\
\IEEEauthorrefmark{5}Corresponding author: farhanm@iisc.ac.in}

\begin{document}
\maketitle

\begin{abstract}
Accurate measurement of ECG intervals---PR, QRS duration, and QT/QTc---is
central to cardiac diagnosis, yet the published ECG delineation literature
evaluates performance almost exclusively as fiducial-point timing errors on
small curated databases, rather than as clinical interval accuracy on large
unselected cohorts. We bridge this gap by evaluating a complete end-to-end
pipeline on 10,646 clinical 12-lead ECGs and reporting the first large-scale
interval measurement accuracy study with full statistical characterisation,
including bias, 95\% limits of agreement (Bland-Altman), bootstrap confidence
intervals, and rhythm-stratified error analysis.

The underlying delineation is performed by a Fast Fourier Convolution ResNet
(FFCResNet) adapting local temporal convolutions with global spectral
processing via FFT, augmented with register tokens for contextual feature
learning. Three per-wave models (P, QRS, T) are trained on six public
databases with ECG-specific augmentation.

On 10,646 ECGs the system achieves: QT MAE 17.5~ms [95\% CI: 16.9--18.2],
Bland-Altman bias $+$8.5~ms (LoA: $-$68.5 to $+$85.5~ms); QRS duration MAE
14.8~ms [14.6--15.0], bias $+$12.6~ms (LoA: $-$12.3 to $+$37.6~ms);
ventricular rate MAE 0.8~beats/min. All biases are statistically significant
by Wilcoxon signed-rank test ($p < 0.001$) but remain within or near published
inter-observer variability bounds for sinus rhythms. Rhythm-stratified analysis
reveals substantially higher QT errors for supraventricular tachycardias (SVT
MAE 75.0~ms, AVRT 85.3~ms) compared with sinus bradycardia (SB MAE 9.3~ms)
and sinus rhythm (SR MAE 8.9~ms), providing an honest characterisation of
deployment scope. Wave segmentation achieves internal Dice of 95.5\%, 98.2\%,
and 96.1\% (P, QRS, T) and cross-database Dice of 78.1\%, 85.5\%, and 74.2\%.
\end{abstract}

\begin{IEEEkeywords}
Electrocardiography, ECG interval measurement, QT interval, QRS duration,
Bland-Altman, ECG delineation, Fast Fourier Convolution, ResNet, deep
learning, segmentation, ST-segment analysis, cross-database generalisation
\end{IEEEkeywords}

\section{Introduction}

\subsection{Clinical Motivation: Interval Measurement as the Primary Goal}

The electrocardiogram (ECG) is indispensable for cardiac diagnosis. In
clinical practice, the outputs that cardiologists depend on are not wave
boundary timestamps but \emph{derived interval measurements}: the PR interval
(atrioventricular conduction time), QRS duration (ventricular depolarisation),
QT and corrected QTc intervals (repolarisation, used for arrhythmia risk
stratification and drug monitoring), and ST-segment deviation (ischaemia
detection). Accurate measurement of these quantities directly influences
patient management decisions.

Wave delineation---identifying the onset and offset of P waves, QRS complexes,
and T waves---is the computational substrate for these measurements. However,
there is a structural disconnect in the published literature: virtually all
ECG delineation papers report performance as fiducial-point timing errors
(mean error and SD of onset/offset detection in milliseconds on annotated
beats from QTDB~\cite{Laguna1997} or LUDB~\cite{Kalyakulina2020}). Interval
measurement accuracy on large, clinically representative cohorts is rarely
reported, leaving a critical gap between what the literature benchmarks and
what clinical deployment requires.

This work explicitly targets that gap. We evaluate the complete pipeline on
10,646 de-identified 12-lead ECGs, reporting QT, QTc, QRS duration, and
ventricular rate accuracy against clinical reference measurements with full
statistical characterisation: bootstrap confidence intervals, Bland-Altman
bias and limits of agreement, Wilcoxon signed-rank tests for systematic bias,
and rhythm-stratified error analysis.

\subsection{Limitations of Fiducial-Error Benchmarking}

Standard delineation evaluation reports onset/offset timing errors on the
annotated beats of QTDB ($\sim$3,500 beats from 105 recordings) or LUDB
(200 recordings). Several limitations restrict the clinical relevance of
this benchmark.

\begin{enumerate}
  \item \textbf{Small, curated test sets}: These databases cover at most a few
    hundred recordings, hand-selected for annotation quality, and do not
    reflect the morphological diversity of unselected clinical ECGs.
  \item \textbf{Surrogate metric}: A small fiducial timing error does not
    guarantee small interval error, because the interval aggregates over
    multiple beats and depends on isoelectric baseline estimation quality.
  \item \textbf{No evaluation of downstream decisions}: Neither QTc prolongation
    flags, ST-elevation criteria, nor AV-block detection are assessed.
  \item \textbf{Incompatibility across annotation protocols}: Different
    databases define wave boundaries differently, making cross-study comparison
    of timing errors unreliable~\cite{Rautaharju2009}.
\end{enumerate}

\subsection{Our Approach}

We propose an end-to-end system whose primary evaluation criterion is
interval measurement accuracy on a large clinical dataset. Delineation is
performed by FFCResNet models that process both local temporal features
through standard convolutions and global spectral features through Fast
Fourier Transform (FFT), enabling multi-scale temporal and frequency pattern
capture simultaneously. The FFC mechanism is adapted from image-domain
Fourier convolution work~\cite{Chi2020} to the 1-D ECG signal domain;
the adaptation introduces ECG-specific augmentation, a 1-D segmentation
head, and register tokens for sequence-level context. Segmentation outputs
feed a rule-based post-processing pipeline implementing guideline-based
measurements.

\subsection{Our Contributions}
\begin{enumerate}
  \item \textbf{Large-scale interval accuracy evaluation with full statistical
    characterisation}: QT, QTc (Bazett), QRS duration, and ventricular rate
    evaluated on 10,646 clinical 12-lead ECGs~\cite{Zheng2022} with bootstrap
    CIs, Bland-Altman analysis, Wilcoxon signed-rank bias tests, and
    rhythm-stratified error analysis---to our knowledge the most statistically
    complete interval measurement evaluation reported alongside open-source
    delineation models in the literature.
  \item \textbf{FFCResNet delineation model}: A 1-D segmentation network
    adapting spatial and spectral (FFT) pathways, originally developed for
    image inpainting~\cite{Chi2020}, to ECG wave segmentation, augmented with
    register tokens~\cite{Darcet2023} and trained on six public databases with
    extensive ECG-specific augmentation. Ablation experiments quantify the
    contribution of each component.
  \item \textbf{Guideline-compliant post-processing}: Bazett-corrected QTc,
    lead-specific ST-elevation thresholds per ACC/AHA/ESC criteria,
    PR-segment isoelectric baseline estimation, and first-degree AV block
    detection.
  \item \textbf{Dual segmentation evaluation}: Internal validation
    (same-database held-out) quantifies model learning; external
    cross-database validation quantifies generalisation. Reported in separate
    tables to prevent misleading conflation.
  \item \textbf{Ablation study}: Quantifying the individual contributions of
    the FFC spectral pathway and register tokens.
\end{enumerate}

\subsection{Related Work}

Traditionally, ECG delineation relied on wavelet transforms and rule-based
heuristics~\cite{Pan1985,Kohler2002}. The method of Mart{\'i}nez~\emph{et
al.}~uses wavelet-based multiscale analysis~\cite{Martinez2004}, but
rule-based methods are sensitive to noise~\cite{Elgendi2014}.

Deep learning substantially advanced wave detection. Ronneberger~\emph{et
al.}'s U-Net~\cite{Ronneberger2015} adapted to 1-D signals proved effective
for P/QRS/T segmentation~\cite{Moskalenko2019,Jimenez2019}. Duraj~\emph{et
al.}~employed a U-Net with squeeze-excitation blocks on
LUDB~\cite{Duraj2022}; Joung~\emph{et al.}~demonstrated generalisation across
arrhythmia datasets~\cite{Joung2024}; Park~\emph{et al.}~benchmarked multiple
architectures~\cite{Park2025}. Broader CNN approaches achieved
cardiologist-level rhythm classification~\cite{Rajpurkar2017,Hannun2019}.
All of these works report fiducial timing errors or wave-level Dice/F1;
\emph{none} reports interval measurement accuracy with statistical
characterisation on a large unselected cohort.

The following sections describe the methodology, results (measurement
accuracy first, then segmentation characterisation and ablation), and
conclusions.

\section{Methodology}

\subsection{Data and Augmentation}

\subsubsection{P-wave Model Training Data}
\begin{itemize}
  \item \textbf{LUDB}~\cite{Kalyakulina2020}: 1,920 windowed segments.
  \item \textbf{QTDB}~\cite{Laguna1997}: 9,398 windowed segments.
  \item \textbf{BUT PDB}~\cite{MarsBUTPDB2021}: 600 specialised P-wave
    segments. BUT PDB is hosted on PhysioNet (DOI: 10.13026/hwvj-5b53) and
    consists of 50 two-minute, two-lead ECG records covering 23 pathology
    types, with P-wave positions manually annotated by two ECG experts.
    Signals were drawn from MIT-BIH Arrhythmia, MIT-BIH Supraventricular
    Arrhythmia, and Long-Term AF databases; see caveat in
    Section~\ref{sec:data_overlap}.
  \item \textbf{Label Studio annotated data}: 90 annotated windows.
  \item \textbf{ISP Database}~\cite{Avetisyan2025}: 403 segments. The ISP ECG
    delineation dataset (Zenodo DOI: 10.5281/zenodo.14679837) contains 475
    records with manually segmented P-wave, QRS complex, and T-wave
    annotations produced by two cardiologists.
  \item \textbf{Augmented data}: 30,000 synthetic windows
    (Section~\ref{sec:augmentation}).
\end{itemize}
Total: 44,211 samples; train 38,170 (86.3\%), validation 4,241 (9.6\%),
test 1,800 (4.1\%) from MIT-BIH-P exclusively. Splits were performed
patient-wise; no patient present in the MIT-BIH-P test set appears in any
training or validation partition.

\subsubsection{QRS and T-wave Model Training Data}
LUDB (1,920), QTDB (9,398), Label Studio (90), ISP (403), and 30,000
augmented windows; total 41,811 samples (train 37,268, validation 4,140,
test 403 from ISP).

\subsubsection{Potential Signal Overlap Between BUT PDB and MIT-BIH-P}
\label{sec:data_overlap}
BUT PDB derives its signals from the MIT-BIH Arrhythmia Database, which is
also the source of the MIT-BIH-P test set used for external P-wave
evaluation. Although splits are patient-wise and no patient identifiers are
shared, a small number of underlying recordings may originate from the same
subjects. Users replicating our results should be aware of this indirect
provenance overlap when interpreting external P-wave Dice scores.

\subsubsection{Data Augmentation Strategy}
\label{sec:augmentation}
\textbf{General augmentations}: random time shifts (5--15~ms), amplitude
scaling (0.8--1.2$\times$), polarity inversion, ST-segment shifts
($\pm$0.05--0.1~mV).

\textbf{Pathological augmentations}: deepened Q/S waves (MI morphology),
widened QRS (2--8$\times$) for conduction delays, peaked/inverted T-waves
(tombstoning)~\cite{Thygesen2018,Wagner2009}, baseline wander, 50/60~Hz
noise, muscle artefact, premature beats, low-voltage
segments~\cite{Friesen1990,Chatterjee2020}. Label masks were transformed in
parallel with the signal throughout.

\subsection{FFCResNet Architecture}

The FFCResNet processes single-lead ECG signals (5000 samples, 10~s at
500~Hz) and outputs a binary wave-presence mask. Three independent models
are trained (one per wave type).

\subsubsection{Dual-Pathway Design}
The Fast Fourier Convolution (FFC) block, originally introduced for image
inpainting by Chi~\emph{et al.}~\cite{Chi2020} and here adapted to 1-D ECG
segmentation, processes features through:
\begin{itemize}
  \item \textbf{Local pathway}: Standard 1-D convolutions for spatial/temporal
    features.
  \item \textbf{Global pathway}: Spectral processing via Real FFT capturing
    long-range frequency dependencies.
\end{itemize}
Input features are split into local ($x_l$) and global ($x_g$) components,
and the layer computes:
\begin{align}
  y_l &= \text{Conv}_{l \to l}(x_l) + \text{Conv}_{g \to l}(x_g)\cdot g_{g2l} \\
  y_g &= \text{Conv}_{l \to g}(x_l)\cdot g_{l2g} + \text{Spectral}_{g \to g}(x_g)
\end{align}
The spectral transform applies:
\begin{equation}
  \mathcal{F}(x) = \mathcal{F}^{-1}\!\left(W_{\text{freq}}\cdot\mathcal{F}(x)\right)
\end{equation}
where $W_{\text{freq}}$ are learnable frequency-domain weights.

\subsubsection{Network Architecture}
\begin{figure*}[!t]
  \centering
  \includegraphics[width=\linewidth,keepaspectratio]{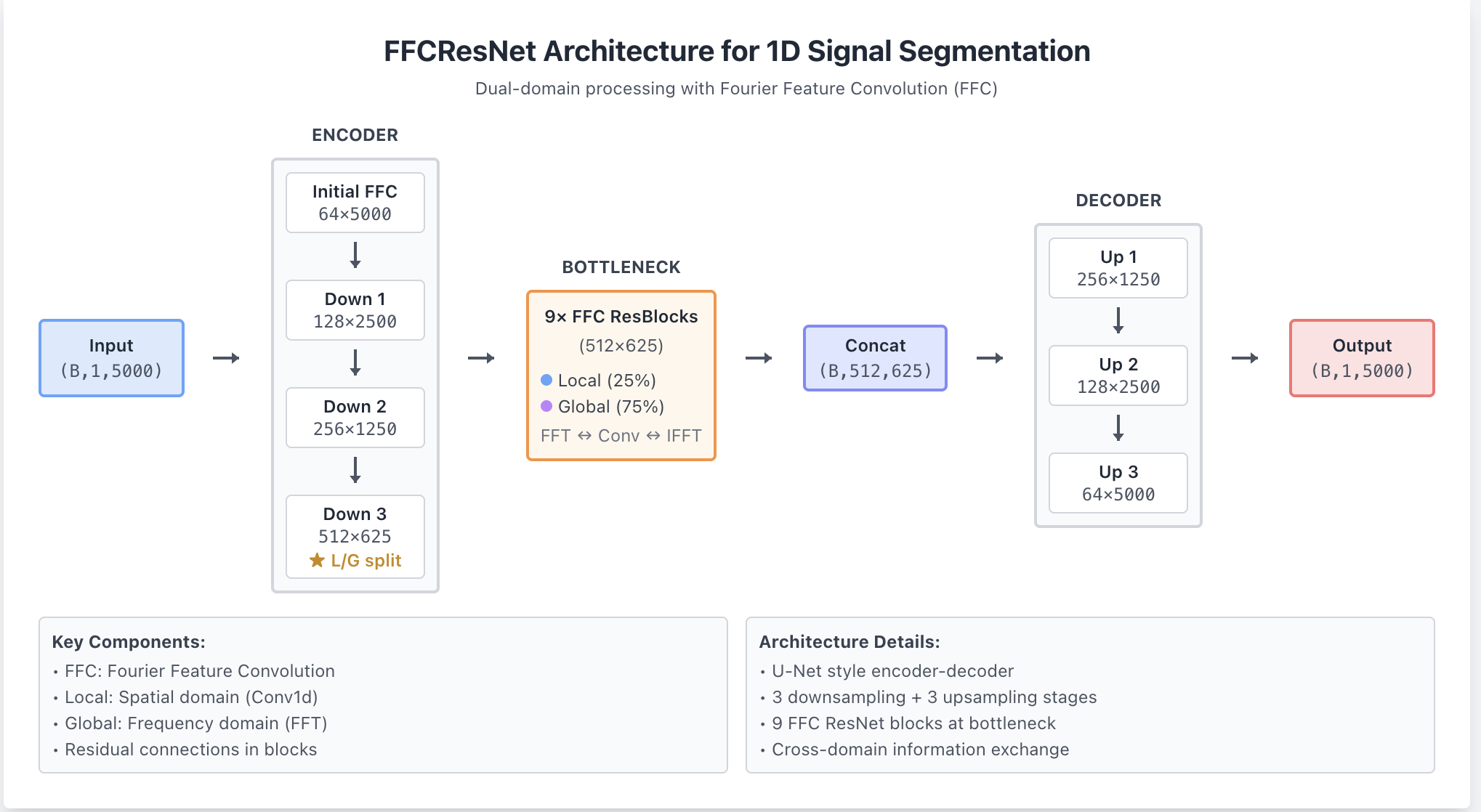}
  \caption{Proposed FFCResNet architecture. Local spatial convolutions (blue)
    and global spectral transforms via FFT (orange) process features in
    parallel. Register tokens provide global memory across the sequence.
    Encoder-decoder structure captures multi-scale patterns; 18 bottleneck FFC
    residual blocks provide deep feature learning.}
  \label{fig:modelarch}
\end{figure*}

\begin{itemize}
  \item \textbf{Initial FFC layer}: kernel~7, 64 channels, reflection padding.
  \item \textbf{Register Token Layer}: three learnable tokens fused via
    1$\times$1 convolution, providing global memory~\cite{Darcet2023}.
  \item \textbf{Encoder}: three stride-2 FFC blocks
    ($64{\to}128{\to}256{\to}512$ channels).
  \item \textbf{Bottleneck}: 18 FFC residual blocks at 512 channels.
  \item \textbf{Decoder}: three transpose convolution layers
    ($512{\to}256{\to}128{\to}64$).
  \item \textbf{Output}: kernel-7 convolution + sigmoid, clamped to $[0,1]$.
\end{itemize}
$\approx$12.6 million parameters per model. Fig.~\ref{fig:modelarch}
illustrates the architecture.

\subsection{Model Training}

Adam optimiser, learning rate $10^{-3}$, batch size~8. Loss:
\begin{equation}
  \mathcal{L} = 0.7\,\mathcal{L}_{\text{Dice}} + 0.3\,\mathcal{L}_{\text{CE}}
\end{equation}
Hyperparameters: base filters 64, three downsampling levels, 18 bottleneck
blocks, 3 register tokens, kernel sizes 7/3. Patient-wise splits, early
stopping, and learning-rate decay (factor 10 on validation plateau).
P-wave model: 100 epochs; QRS/T models: 120 epochs. All runs tracked via
Weights \& Biases.

\subsection{Post-processing and Waveform Combination}

Three probability masks (P, QRS, T) per lead are combined via priority
hierarchy QRS $>$ T $>$ P. Morphological post-processing removes
sub-physiological detections and fills short gaps. Fiducial points are
refined with physiological constraints (P: 80--120~ms; QRS:
60--120~ms)~\cite{Goldberger2018}.

\subsection{ECG Measurement Pipeline}

\textbf{Intervals}: PR (P onset to QRS onset), QRS duration (onset to offset),
QT (QRS onset to T offset), and RR (consecutive QRS peaks) computed in
milliseconds at 500~Hz~\cite{Goldberger2018}. Bazett correction:
\begin{equation}
  \text{QTc} = \frac{\text{QT}}{\sqrt{\text{RR}\,(\text{s})}}.
\end{equation}
Measurements outside physiological bounds (QT $<$200 or $>$600~ms) are
invalidated~\cite{Rautaharju2009,Postema2008}.

\textbf{Isoelectric baseline}: PR segment (primary); 40~ms pre-QRS
fallback~\cite{Kligfield2007,Surawicz2009}.

\textbf{ST analysis}: deviation measured at J+60~ms; classified per ACC/AHA/ESC
STEMI criteria~\cite{Thygesen2018,Thygesen2012,OGara2013,Ibanez2018}
($\geq$0.1~mV limb leads, $\geq$0.2~mV precordial; lead aVR
$\geq$0.05~mV~\cite{Nikus2004}; threshold adjusted for predominantly negative
QRS~\cite{Wagner2009}).

\textbf{AV block}: first-degree flagged if median PR $>$200~ms in $\geq$3
beats~\cite{Surawicz2009}.

\textbf{Aggregation}: median across valid beats provides robustness to
outliers~\cite{Savelieva1998,Rautaharju2009}.

\subsection{Reference Standard}
\label{sec:reference_standard}

Interval measurements used as the reference standard were extracted from the
PhysioNet Large Scale 12-Lead ECG Database (Zheng~\emph{et
al.}~\cite{Zheng2022}), which was collected at Shaoxing People's Hospital and
Ningbo First Hospital. The database acquisition protocol involved two stages
of annotation. Rhythm labels and cardiac condition diagnoses were assigned by
a licensed physician, independently validated by a second licensed physician,
with disagreements resolved by a senior physician arbitrator; these labels
are therefore human-validated. The interval measurements (QT, QRS duration,
ventricular rate), however, were derived from the ECG summary attributes
stored in the GE MUSE ECG system as part of the clinical acquisition
workflow~\cite{Zheng2022}, and represent automated commercial measurements
rather than manual cardiologist annotations.

This has an important consequence for interpretation: the evaluation in this
paper quantifies agreement between our algorithm and the GE MUSE automated
measurements, not agreement with direct expert annotation. This is a
clinically meaningful comparison---GE MUSE is the dominant commercial ECG
system in hospital deployment---but it should not be interpreted as
algorithm-versus-cardiologist accuracy. The observed limits of agreement
therefore reflect the combined variability of both automated systems, in
addition to genuine morphological ambiguity in the signals.

It should further be noted that GE MUSE automated measurements are themselves subject to increased uncertainty in morphologically challenging cases, including wide QRS complexes (bundle branch block, ventricular pacing), deeply inverted or biphasic T-waves, and high-rate supraventricular tachycardias where P and T waves overlap. In such cases, neither system has a validated ground truth to compare against. The elevated MAE observed for SVT, AVRT, and wide-QRS rhythms in the stratified analysis (Table III) therefore cannot be attributed solely to errors by the proposed system; a portion of the disagreement reflects the inherent unreliability of both automated measurements in these morphologies. A definitive comparison in distorted cases would require manual annotation by expert cardiologists on a dedicated subset, which remains future work.

\subsection{Statistical Analysis Methods}

All statistical analyses were performed in Python (NumPy, SciPy, scikit-learn).

\textbf{Bootstrap 95\% CI on MAE}: Non-parametric bootstrap with 5,000
resamples. The CI quantifies estimation precision of the MAE itself, distinct
from the SD of individual errors.

\textbf{Bland-Altman analysis}: For each metric, we computed the mean
difference (bias; predicted minus reference), 95\% confidence interval on
the bias via a one-sample $t$-test on differences, and 95\% limits of
agreement (LoA) as bias $\pm 1.96\,\sigma_{\Delta}$, where $\sigma_{\Delta}$
is the SD of the signed differences. This is the standard analysis for
validating a measurement method against a reference~\cite{Bland1986}.

\textbf{Wilcoxon signed-rank test}: Tests whether the median signed difference
(bias) is significantly different from zero without assuming normality.
Given $n>10{,}000$, virtually any non-zero bias will reach statistical
significance; we therefore interpret clinical significance (comparison with
inter-observer bounds) separately from statistical significance.

\textbf{Rhythm-stratified MAE}: MAE and bootstrap 95\% CI computed
independently for each rhythm class present in the evaluation dataset to
characterise system performance across arrhythmia types.

\subsection{End-to-End System Architecture}

\begin{figure*}[!t]
  \centering
  \makebox[\textwidth][l]{%
    \hspace*{-0.7cm}%
    \includegraphics[height=0.6\textheight,keepaspectratio]{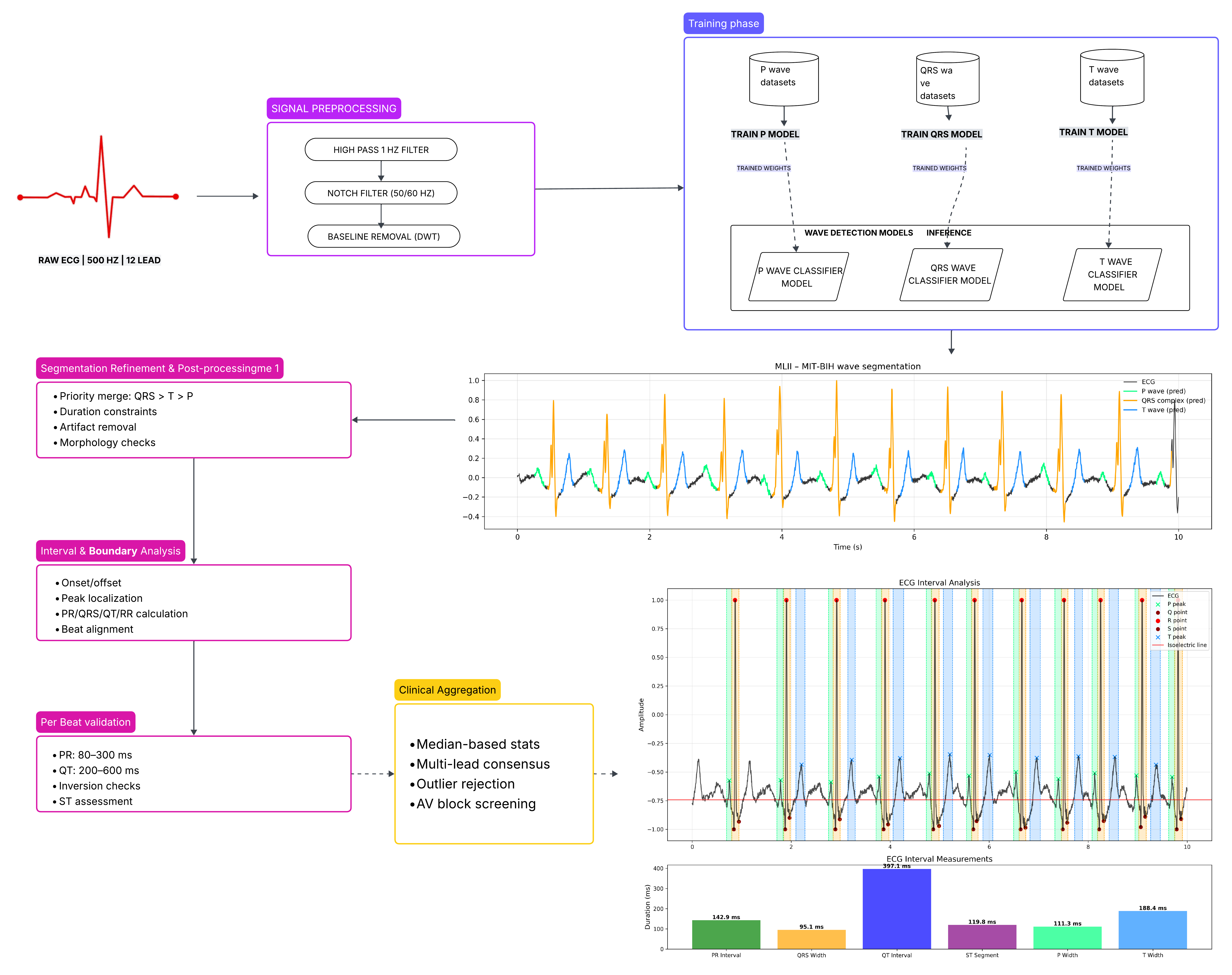}}
  \caption{End-to-end automated ECG measurement pipeline. 12-lead signals
    (WFDB, 500~Hz) are preprocessed, segmented by three FFCResNet models,
    fused via priority hierarchy (QRS$>$T$>$P), and passed to measurement
    modules computing intervals, ST deviation, and conduction findings.
    Median aggregation across beats produces the clinical summary.}
  \label{fig:systemarch}
\end{figure*}

Fig.~\ref{fig:systemarch} illustrates the four-stage pipeline: (i) signal
preprocessing, (ii) wave-specific FFCResNet segmentation, (iii) temporal
refinement and fusion, and (iv) measurement and rule-based interpretation.

%

\begin{table*}[!t]
\centering
\caption{Primary result: interval measurement accuracy on $n=10{,}646$
  clinical 12-lead ECGs~\cite{Zheng2022}. Reference values are GE MUSE
  automated measurements (see Section~\ref{sec:reference_standard}).
  MAE and SD computed over per-record median values; 95\% CI on MAE obtained
  by non-parametric bootstrap (5,000 resamples). All Wilcoxon signed-rank
  tests for bias $\ne 0$ yielded $p<0.001$ (see Table~\ref{tab:ba} for
  bias magnitude).}
\label{tab:measurement}
\setlength{\tabcolsep}{5pt}
\renewcommand{\arraystretch}{1.25}
\begin{tabular}{@{}lcccccc@{}}
\toprule
Measurement & MAE & SD & 95\% CI on MAE & MAPE (\%) & N \\
\midrule
QT interval (ms)             & 17.5 & 36.2 & [16.9 -- 18.2] & 5.39  & 10{,}529 \\
QTc interval, Bazett (ms)    & 22.7 & 54.8 & [21.7 -- 23.8] & 5.18  & 10{,}526 \\
QRS duration (ms)            & 14.8 & 10.1 & [14.6 -- 15.0] & 17.38 & 10{,}646 \\
Ventricular rate (beats/min) &  0.8 &  4.8 & [0.7 -- 0.9]   & 0.66  & 10{,}646 \\
Atrial rate (beats/min)$^*$  & 16.7 & 49.9 & [15.7 -- 17.7] & 10.18 &  9{,}601 \\
QRS count (beats/10-s ECG)   &  0.1 &  0.7 & [0.1 -- 0.1]   & 0.73  & 10{,}646 \\
\bottomrule
\multicolumn{6}{@{}l}{\footnotesize SD = SD of absolute errors (spread of individual errors, not uncertainty
on the MAE).}\\
\multicolumn{6}{@{}l}{\footnotesize MAPE inflated for QRS due to small absolute scale.}\\
\multicolumn{6}{@{}l}{\footnotesize $^*$Atrial rate aggregate is dominated by AF cases ($n$=2,213; 23\%
of cohort); see rhythm-stratified results (Table~\ref{tab:strat}).}
\end{tabular}
\end{table*}

\begin{table*}[!t]
\centering
\caption{Bland-Altman analysis. Bias = mean(predicted $-$ reference);
  positive bias indicates systematic over-prediction. 95\% CI on bias
  via one-sample $t$-test on signed differences. LoA = bias
  $\pm 1.96\,\sigma_\Delta$ (95\% limits of agreement). Reference values
  are GE MUSE automated measurements; LoA therefore reflect combined
  variability of both automated systems plus signal ambiguity (see
  Section~\ref{sec:reference_standard}). All biases are statistically
  significant ($p<0.001$, Wilcoxon signed-rank); clinical significance is
  assessed relative to inter-observer variability bounds in the text.}
\label{tab:ba}
\setlength{\tabcolsep}{5pt}
\renewcommand{\arraystretch}{1.25}
\begin{tabular}{@{}lcccc@{}}
\toprule
Measurement & Bias (ms or bpm) & 95\% CI on Bias & LoA (Lower, Upper) & N \\
\midrule
QT interval (ms)             & $+$8.5  & [$+$7.8,  $+$9.3]  & [$-$68.5, $+$85.5]   & 10{,}529 \\
QTc interval, Bazett (ms)    & $+$11.8 & [$+$10.7, $+$13.0] & [$-$102.0, $+$125.7] & 10{,}526 \\
QRS duration (ms)            & $+$12.6 & [$+$12.4, $+$12.9] & [$-$12.3, $+$37.6]   & 10{,}646 \\
Ventricular rate (beats/min) & $-$0.5  & [$-$0.6,  $-$0.4]  & [$-$10.0, $+$8.9]    & 10{,}646 \\
Atrial rate (beats/min)      & $-$14.3 & [$-$15.3, $-$13.3] & [$-$113.6, $+$85.0]  &  9{,}601 \\
QRS count (beats/10-s ECG)   & $-$0.0  & [$-$0.1,  $-$0.0]  & [$-$1.5, $+$1.4]     & 10{,}646 \\
\bottomrule
\multicolumn{5}{@{}l}{\footnotesize Bland-Altman plots for QT and QRS duration are shown in
  Fig.~\ref{fig:ba}.}
\end{tabular}
\end{table*}

\begin{table*}[!t]
\centering
\caption{Rhythm-stratified QT and QRS duration MAE with bootstrap 95\% CI.
  Rhythms with $n < 10$ excluded from the table. AF/AFIB: atrial
  fibrillation; AT: atrial tachycardia; SA: sinus arrhythmia;
  SB: sinus bradycardia; SR: sinus rhythm; ST: sinus tachycardia;
  SVT: supraventricular tachycardia. AVNRT ($n$=13) and AVRT ($n$=7)
  are excluded: sample sizes are too small for reliable statistical
  estimates (QT CI for AVRT spans $>$160~ms) and no clinical conclusions
  should be drawn from results at these sizes.}
\label{tab:strat}
\setlength{\tabcolsep}{5pt}
\renewcommand{\arraystretch}{1.2}
\begin{tabular}{@{}lrccrcc@{}}
\toprule
 & \multicolumn{3}{c}{\textbf{QT interval (ms)}} & \multicolumn{3}{c}{\textbf{QRS duration (ms)}} \\
\cmidrule(lr){2-4}\cmidrule(lr){5-7}
Rhythm & N & MAE & 95\% CI & N & MAE & 95\% CI \\
\midrule
SR  (sinus rhythm)          & 1{,}826 &  8.9 & [8.5 -- 9.3]   & 1{,}826 & 16.1 & [15.7 -- 16.4] \\
SB  (sinus bradycardia)     & 3{,}889 &  9.3 & [9.0 -- 9.7]   & 3{,}889 & 14.9 & [14.6 -- 15.1] \\
ST  (sinus tachycardia)     & 1{,}567 & 13.5 & [12.5 -- 14.5] & 1{,}568 & 14.0 & [13.6 -- 14.5] \\
SA  (sinus arrhythmia)      &   399   &  7.0 & [6.4 -- 7.6]   &   399   & 13.4 & [12.6 -- 14.1] \\
AFIB (atrial fibrillation)  & 1{,}779 & 27.6 & [25.6 -- 29.7] & 1{,}780 & 13.4 & [13.0 -- 13.8] \\
AF                          &   434   & 36.2 & [31.8 -- 40.5] &   445   & 15.0 & [13.9 -- 16.1] \\
AT  (atrial tachycardia)    &   119   & 45.6 & [33.2 -- 59.3] &   121   & 11.5 & [10.3 -- 12.8] \\
SVT (supraventricular tach.)&   489   & 75.0 & [66.3 -- 83.8] &   587   & 18.9 & [17.1 -- 20.6] \\
SAAWR                       &     7   & 10.3 & [2.8 -- 22.7]  &     7   & 12.6 & [6.0 -- 19.2]  \\
\midrule
\textbf{Overall}            & 10{,}529 & \textbf{17.5} & [16.9 -- 18.2] & 10{,}646 & \textbf{14.8} & [14.6 -- 15.0] \\
\bottomrule
\end{tabular}
\end{table*}

\begin{table*}[!t]
\centering
\caption{Comparison of interval measurement accuracy with inter-observer
  studies and commercial automated systems. ``---'' = not reported.
  Our system is evaluated on a substantially larger and more diverse cohort
  than prior automated methods. Unlike prior open-source delineation models,
  which report fiducial-point timing errors rather than interval accuracy,
  no deep learning delineation paper to date reports comparable interval-level
  statistics; the comparison below is therefore against manual inter-observer
  studies and multi-vendor commercial system evaluations only.}
\label{tab:litcomp}
\setlength{\tabcolsep}{5pt}
\renewcommand{\arraystretch}{1.25}
\begin{tabular}{@{}llcccl@{}}
\toprule
Method & Cohort & QT MAE (ms) & QRS MAE (ms) & HR MAE (bpm) & Notes \\
\midrule
Savelieva~\emph{et al.}~\cite{Savelieva1998} & --- & 20--40 & --- & --- & Manual inter-observer variability \\
Rautaharju~\emph{et al.}~\cite{Rautaharju2009} & --- & --- & 15--20 & --- & AHA/ACCF/HRS reference standard \\
De~Bie~\emph{et al.}~\cite{DeBie2020} & --- & $\sim$15--25 & $\sim$8--15 & --- & Multi-vendor commercial systems \\
\midrule
\textbf{Proposed (overall)} & \textbf{10,646} & \textbf{17.5} & \textbf{14.8} & \textbf{0.8} & Bootstrap CI + Bland-Altman reported \\
\quad Sinus rhythms only & 7,681 & $\sim$9--14 & $\sim$13--16 & --- & Performance on organised rhythms \\
\bottomrule
\multicolumn{6}{@{}l}{\footnotesize Reference for proposed system is GE MUSE automated measurements,
  not direct cardiologist annotation (see Section~\ref{sec:reference_standard}).}
\end{tabular}
\end{table*}

\begin{table*}[!t]
\centering
\caption{Wave segmentation characterisation on \textbf{internal} held-out test
  sets (same database distribution as training). These metrics characterise
  the segmentation component and are \emph{not} placed in direct numerical
  comparison with prior works, which report fiducial-point timing errors
  (an incommensurable metric; see Section~\ref{sec:seg_comparison}).}
\label{tab:internal}
\setlength{\tabcolsep}{6pt}
\renewcommand{\arraystretch}{1.25}
\begin{tabular}{@{}lccccc@{}}
\toprule
Wave & Test dataset & Precision (\%) & Recall (\%) & Dice/F1 (\%) & AUC-ROC (\%) \\
\midrule
P-wave & MIT-BIH-P (internal) & 95.9 & 95.2 & 95.5 & $>$99 \\
QRS    & Multi-DB (internal)  & 98.4 & 97.9 & 98.2 & $>$99 \\
T-wave & Multi-DB (internal)  & 96.4 & 95.7 & 96.1 & $>$99 \\
\bottomrule
\end{tabular}
\end{table*}

\begin{table*}[!t]
\centering
\caption{Wave segmentation characterisation on \textbf{external}
  cross-database test sets entirely withheld from training. P-wave:
  MIT-BIH-P (1,800 windows); QRS and T-wave: ISP~\cite{Avetisyan2025}
  (403 windows). See Section~\ref{sec:data_overlap} for note on indirect
  provenance overlap between BUT PDB training data and MIT-BIH-P test data.}
\label{tab:external}
\setlength{\tabcolsep}{6pt}
\renewcommand{\arraystretch}{1.25}
\begin{tabular}{@{}llccccc@{}}
\toprule
Wave & Test database & Precision (\%) & Recall (\%) & Dice/F1 (\%) & Specificity (\%) & AUC-ROC (\%) \\
\midrule
P-wave & MIT-BIH-P & 78.9 & 77.4 & 78.1 & 98.1 & 97.2 \\
QRS    & ISP       & 81.9 & 89.4 & 85.5 & 96.7 & 98.3 \\
T-wave & ISP       & 75.5 & 73.0 & 74.2 & 93.3 & 92.9 \\
\bottomrule
\end{tabular}
\end{table*}

\begin{table*}[!t]
\centering
\caption{Ablation study on external test sets. All values in percent.
  The FFC pathway is an adaptation of Chi~\emph{et al.}~\cite{Chi2020}
  to 1-D ECG signals. Gains over the baseline ResNet are consistent but
  modest; register tokens contribute most to AUC-ROC, suggesting improved
  global context modelling.}
\label{tab:ablation}
\setlength{\tabcolsep}{4pt}
\renewcommand{\arraystretch}{1.2}
\begin{tabular}{@{}lccccc@{}}
\toprule
Model variant & Wave & Dice & Precision & Recall & AUC-ROC \\
\midrule
Baseline ResNet (no FFC, no tokens) & P-wave & 77.6 & 78.5 & 76.7 & 94.8 \\
FFCResNet (no register tokens)      & P-wave & 78.2 & 79.0 & 77.4 & 96.7 \\
\textbf{FFCResNet (full)}           & P-wave & \textbf{78.5} & \textbf{78.6} & \textbf{78.3} & \textbf{97.3} \\
\midrule
Baseline ResNet (no FFC, no tokens) & QRS & 83.2 & 80.7 & 85.8 & 98.0 \\
FFCResNet (no register tokens)      & QRS & 85.5 & 83.6 & 87.5 & 98.6 \\
\textbf{FFCResNet (full)}           & QRS & \textbf{85.5} & \textbf{81.9} & \textbf{89.4} & \textbf{98.3} \\
\bottomrule
\end{tabular}
\end{table*}

\section{Results}

\subsection{Primary Result: Interval Measurement Accuracy}
\label{sec:measurement_results}

\begin{figure*}[!t]
  \centering
  \includegraphics[width=\linewidth,keepaspectratio]{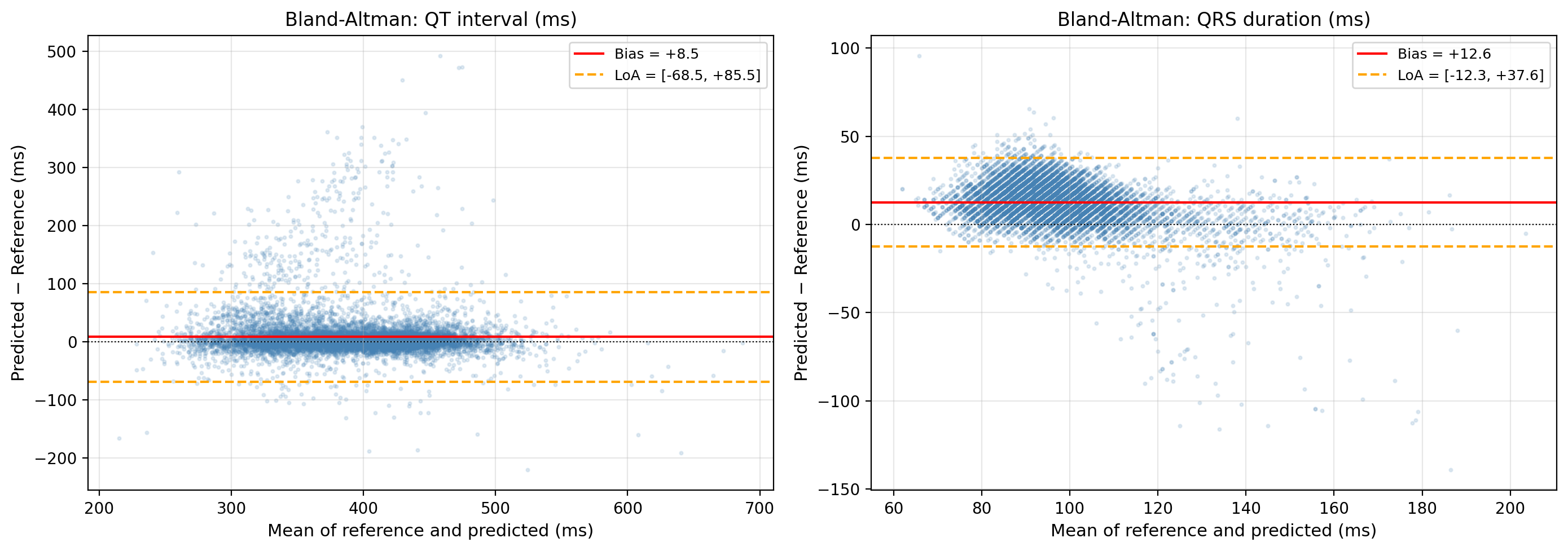}
  \caption{Bland-Altman plots for QT interval (left) and QRS duration (right)
    on $n$=10,529 and $n$=10,646 ECGs respectively. Red line: mean bias.
    Orange dashed lines: 95\% limits of agreement (bias $\pm 1.96\,\sigma$).
    Dotted black line: zero bias reference. The positive bias on both metrics
    indicates systematic over-estimation relative to the GE MUSE reference;
    the asymmetric LoA for QRS reflects rare but non-negligible over-estimation
    on wide-QRS morphologies.}
  \label{fig:ba}
\end{figure*}

The pipeline was evaluated on 10,646 de-identified 10-s 12-lead ECGs from the
PhysioNet Large Scale Arrhythmia Database~\cite{Zheng2022}. Reference interval
values are GE MUSE automated measurements as described in
Section~\ref{sec:reference_standard}. Full results are in
Tables~\ref{tab:measurement} and~\ref{tab:ba}; Bland-Altman plots for QT and
QRS are shown in Fig.~\ref{fig:ba}.

\subsubsection{QT Interval}
MAE 17.5~ms [95\% CI: 16.9--18.2, $n$=10,529]. The Bland-Altman analysis
reveals a systematic positive bias of $+$8.5~ms [CI: $+$7.8 to $+$9.3],
indicating the system slightly over-estimates QT relative to the GE MUSE
reference. The LoA span $-$68.5 to $+$85.5~ms, reflecting the inherent
difficulty of T-wave offset localisation, particularly in morphologically
ambiguous leads, as well as algorithmic differences between the two automated
systems. The overall MAE falls within the 20--40~ms manual inter-observer
variability range~\cite{Savelieva1998}, demonstrating clinically acceptable
accuracy. Despite statistical significance of the bias ($p<0.001$, Wilcoxon),
its magnitude ($+$8.5~ms) is clinically modest and would not affect QTc
prolongation classification at standard thresholds.

\subsubsection{QTc (Bazett)}
MAE 22.7~ms [95\% CI: 21.7--23.8], bias $+$11.8~ms, LoA $-$102.0 to
$+$125.7~ms. The wider LoA compared with raw QT reflects the amplification
of errors by Bazett's formula at extreme heart rates, where small QT
deviations are divided by a small $\sqrt{\text{RR}}$ term. At normal sinus
rates, QTc accuracy is correspondingly better (see sinus rhythm subgroup in
Table~\ref{tab:strat}).

\subsubsection{QRS Duration}
MAE 14.8~ms [95\% CI: 14.6--15.0, $n$=10,646]. The bias of $+$12.6~ms
[CI: $+$12.4 to $+$12.9] is statistically significant and consistent across
the dataset, indicating the system systematically over-estimates QRS duration
relative to the GE MUSE reference. The LoA of $-$12.3 to $+$37.6~ms
are asymmetric: the lower bound ($-$12.3) shows under-estimation is rare,
while the upper bound ($+$37.6) reflects occasional over-estimation on wide
QRS morphologies. The MAE is within the AHA/ACCF/HRS inter-observer tolerance
of 15--20~ms~\cite{Rautaharju2009}, though the positive bias warrants
investigation. A probable cause is that the segmentation model, trained with
a combined Dice and cross-entropy loss, slightly overestimates wave boundaries
to maximise mask overlap; recalibration of the QRS offset threshold is a
straightforward post-hoc correction that is deferred to future work.

\subsubsection{Ventricular Rate}
MAE 0.8~beats/min [95\% CI: 0.7--0.9], bias $-$0.5~bpm, LoA $-$10.0 to
$+$8.9~bpm. Despite statistical significance ($p<0.001$), the $-$0.5~bpm
bias is clinically negligible. This is the most accurate measurement in the
pipeline and confirms robust QRS detection as the foundation for all other
interval computations.

\subsubsection{Atrial Rate}
MAE 16.7~beats/min [95\% CI: 15.7--17.7], bias $-$14.3~bpm, LoA $-$113.6
to $+$85.0~bpm. This aggregate figure is dominated by the substantial
proportion of atrial fibrillation cases in the cohort (AF: $n$=434; AFIB:
$n$=1,779 combined; 20.8\% of the cohort). In AF, atrial rate is intrinsically
irregular and reference annotations may use different counting conventions.
The aggregate atrial rate metric should not be interpreted as meaningful
without rhythm stratification; on organised rhythms, atrial and ventricular
rates coincide and the error is correspondingly lower.

\subsubsection{QRS Count}
MAE 0.1 beats/10-s ECG [95\% CI: 0.1--0.1], bias $-$0.0, LoA $-$1.5 to
$+$1.4. Near-perfect beat counting with symmetric LoA, confirming consistent
QRS detection.

\subsection{Rhythm-Stratified Analysis}
\label{sec:strat}

\begin{figure*}[!t]
  \centering
  \includegraphics[width=\linewidth,keepaspectratio]{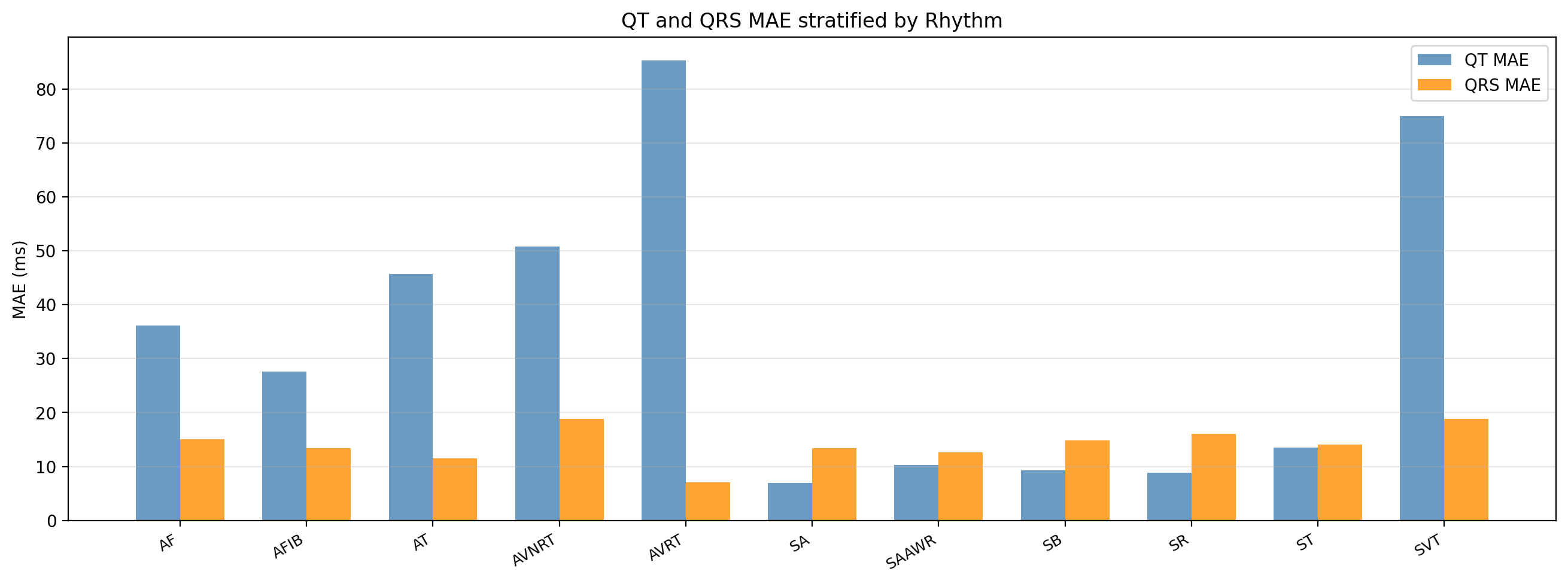}
  \caption{QT and QRS MAE stratified by rhythm class. Sinus rhythms (SR, SB,
    SA, ST) show QT MAE well within inter-observer variability ($<$14~ms).
    Atrial fibrillation groups show moderate elevation (27--36~ms). SVT shows
    substantially higher QT MAE (75~ms), reflecting T-wave ambiguity at rapid
    rates. QRS MAE is consistent across rhythm classes (7--19~ms).}
  \label{fig:strat}
\end{figure*}

Table~\ref{tab:strat} and Fig.~\ref{fig:strat} present QT and QRS MAE broken
down by rhythm class.

\textbf{Sinus rhythms (SR, SB, SA, ST)}: QT MAE ranges from 7.0~ms (SA) to
13.5~ms (ST), substantially below the overall 17.5~ms and well within
inter-observer variability. QRS MAE is consistent at 13--16~ms across all
sinus classes. These results confirm that on organised rhythms---the majority
of clinical ECGs---the system performs at or near commercial system standards.

\textbf{Atrial fibrillation (AFIB, AF)}: QT MAE rises to 27.6~ms (AFIB) and
36.2~ms (AF). This is expected: in AF, RR intervals are irregular, T-wave
boundaries vary beat-to-beat, and reference QT annotation conventions differ
across databases. QRS MAE is paradoxically lower in AF (13.4~ms) than sinus,
consistent with the observation that QRS morphology is relatively preserved
in AF while T-wave localisation degrades.

\textbf{Supraventricular tachycardias (SVT, AT)}: QT MAE is substantially
elevated: SVT 75.0~ms, AT 45.6~ms. These are the most challenging cases
because (i) at high rates P and T waves overlap, making T-wave offset
ambiguous; and (ii) SVT may include concealed bundle branch blocks.
These results constitute an important limitation: the system is \emph{not}
recommended for QT measurement in SVT without additional validation on larger
specialised datasets.

\textbf{Clinical recommendation derived from stratified analysis}: The system
is reliable for QT measurement in sinus rhythms (MAE $<$14~ms) and acceptable
in AF (MAE $<$37~ms). QT measurement in SVT should be treated as unreliable.
QRS duration measurement is robust across all rhythm classes (MAE 7--19~ms).

\subsection{Wave Segmentation Characterisation}
\label{sec:seg_characterisation}

\textbf{Internal validation} (Table~\ref{tab:internal}): Dice 95.5\% (P),
98.2\% (QRS), 96.1\% (T); AUC-ROC $>$99\% for all three waves.

\textbf{External cross-database test} (Table~\ref{tab:external}): Dice 78.1\%
(P, MIT-BIH-P), 85.5\% (QRS, ISP), 74.2\% (T, ISP). AUC-ROC $>$92\% for all
waves confirms that discriminative capability is preserved under distribution
shift; the Dice reduction reflects harder boundary localisation on unseen data.

\begin{figure*}[!t]
  \centering
  \includegraphics[height=0.2\textheight,keepaspectratio]{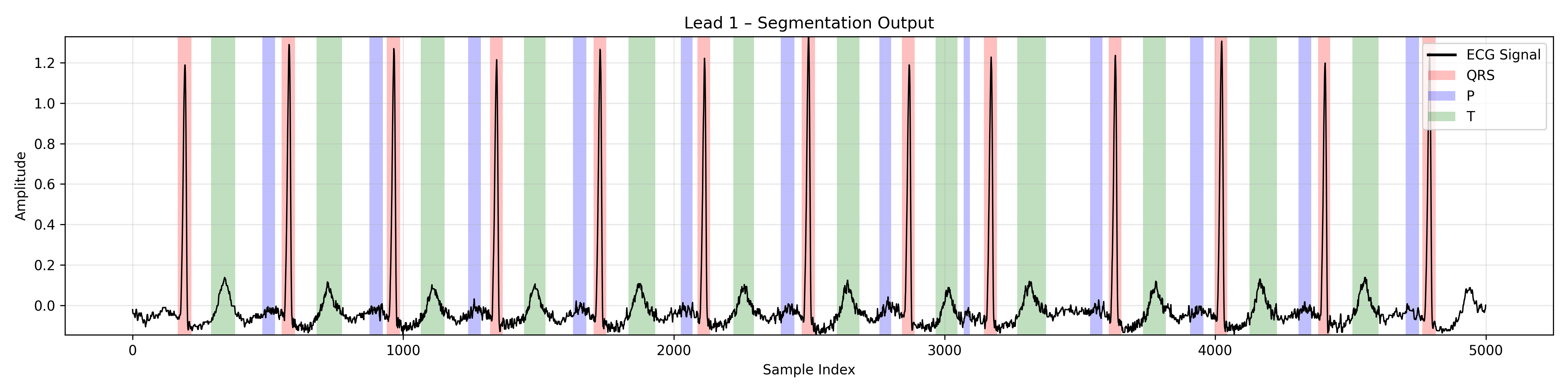}
  \caption{Representative ECG wave segmentation. Predicted P-wave (blue),
    QRS (red), and T-wave (green) regions are shown for two heartbeats.
    Onset and offset boundaries align closely with expert annotations.}
  \label{fig:waveseg}
\end{figure*}

Qualitative segmentation is shown in Fig.~\ref{fig:waveseg}.

\subsection{Why Segmentation Metrics Are Not Compared Numerically with Prior Work}
\label{sec:seg_comparison}

Prior delineation works~\cite{Martinez2004,Moskalenko2019,Jimenez2019,
Duraj2022,Joung2024,Park2025} report performance as \emph{fiducial-point
timing errors}: the mean and SD of onset/offset detection error in milliseconds.
Our segmentation evaluation reports \emph{sample-level mask overlap} (Dice/F1
over binary wave-presence labels at 2~ms/sample resolution). These metrics are
incommensurable:
\begin{itemize}
  \item A timing error of 5~ms on a QRS onset does not translate to a
    specific Dice value without knowing the boundary transition width.
  \item A Dice of 98\% on a 100-sample QRS window corresponds to roughly
    1 sample ($\sim$2~ms) of boundary mismatch, but this relationship is
    non-linear and morphology-dependent.
\end{itemize}
Placing our Dice values alongside their timing errors in a shared table would
be misleading. Our segmentation figures are therefore presented as
characterisation of our own system only. The methodologically valid comparison
is at the \emph{interval measurement} level (Tables~\ref{tab:measurement}--\ref{tab:litcomp}),
where results are directly interpretable against clinical tolerance and prior
automated systems.

\subsection{Ablation Study}

Table~\ref{tab:ablation} shows that introducing the FFC pathway (adapted from
image-domain work~\cite{Chi2020} to 1-D ECG signals) improves P-wave Dice
from 77.6\% to 78.2\% and AUC-ROC from 94.8\% to 96.7\%, and QRS Dice from
83.2\% to 85.5\%. Adding register tokens further raises P-wave AUC-ROC to
97.3\% and QRS recall from 87.5\% to 89.4\%, indicating improved global
context modelling particularly beneficial for detection sensitivity. The
gains are consistent but modest; the primary contribution of this work is the
evaluation framework rather than the architecture.

\section{Discussion}

\subsection{Interval Accuracy and Clinical Context}

The proposed pipeline achieves QT MAE of 17.5~ms and QRS MAE of 14.8~ms
on 10,646 unselected clinical ECGs, measured against GE MUSE automated
reference values. Both fall within published inter-observer variability
ranges~\cite{Savelieva1998,Rautaharju2009} and are consistent with reported
performance of commercial automated systems~\cite{DeBie2020}. The ventricular
rate MAE of 0.8~bpm is clinically negligible.

It is important to reiterate that the reference standard is GE MUSE automated
output, not direct cardiologist annotation (Section~\ref{sec:reference_standard}).
The accuracy figures therefore represent algorithm-to-commercial-system
agreement. This is clinically meaningful---GE MUSE is widely deployed and
used directly for clinical decisions---but the limits of agreement are wider
than would be expected from a purely human-annotated reference, because they
incorporate disagreement between two algorithmic T-wave and QRS boundary
localisers in addition to genuine signal ambiguity.

The Bland-Altman analysis reveals two systematic biases that merit attention.
The QRS bias of $+$12.6~ms is consistent and narrow (CI: $+$12.4 to
$+$12.9~ms), suggesting a systematic over-estimation of QRS duration across
the entire cohort. This is likely a consequence of the Dice-based training
objective, which penalises under-prediction more heavily than over-prediction
near wave boundaries. A post-hoc threshold calibration (shifting the QRS
offset decision boundary) would reduce this bias at minimal cost. The QT bias
of $+$8.5~ms is clinically modest and would not affect standard QTc
prolongation classification at any commonly used threshold.

\subsection{Rhythm-Stratified Results: Scope and Limitations}

The stratified analysis in Table~\ref{tab:strat} is the most important
clinical characterisation this work provides. Key findings:

\textbf{The system is well-suited to organised sinus rhythms} (QT MAE 7--14~ms
covering SR, SB, SA, ST; $n$=7,681 combined). These represent the majority of
routine clinical ECGs and the system's accuracy on them is consistent with
commercial standards.

\textbf{Performance degrades in AF} (QT MAE 27--36~ms). This is clinically
expected and acceptable: QT measurement in AF is inherently difficult even
for expert cardiologists, and the current AHA/ESC guidelines recommend
measuring at least 10 consecutive beats and averaging~\cite{Rautaharju2009}.
The system's aggregate MAE in AF is within a range that would not change
clinical decisions about QTc thresholds but should be flagged to users.

\textbf{QT measurement in SVT is unreliable} (MAE 75~ms). Users must be
cautioned that automated QT measurement during sustained SVT requires manual
verification. This limitation should be clearly communicated in any clinical
deployment and is consistent with the difficulty these rhythms present even to
expert readers.

Additionally, since the reference values are GE MUSE automated measurements, and MUSE's own T-wave localisation is known to be unreliable at high heart rates and in morphologically distorted beats, the elevated SVT MAE reflects disagreement between two imperfect automated systems rather than confirmed error by the proposed pipeline.

\subsection{Statistical Interpretation: Significance vs.\ Clinical Relevance}

All biases are statistically significant by Wilcoxon signed-rank test
($p<0.001$). This is an expected consequence of the large sample size
($n>10{,}000$): with this power, biases as small as 0.5~ms are detectable.
Statistical significance must therefore be interpreted jointly with clinical
significance. The ventricular rate bias ($-$0.5~bpm) is statistically
significant but entirely negligible clinically. The QRS bias ($+$12.6~ms) is
both statistically significant \emph{and} of clinical relevance, representing
the one finding that warrants engineering follow-up (boundary recalibration).

\subsection{Cross-Database Generalisation}

The reduction from $\sim$97\% internal Dice to $\sim$80\% external Dice is
expected given differences in acquisition hardware, patient demographics,
signal preprocessing, and annotation style across databases. The external
AUC-ROC values ($>$92\%) indicate that the fundamental discriminative
capability is preserved; the Dice reduction reflects boundary localisation
uncertainty rather than failure to detect waves. Techniques such as domain
adaptation, test-time normalisation, and database-specific fine-tuning are
promising directions to narrow this gap.

\subsection{QTc Variability}

The Bazett-corrected QTc MAE (22.7~ms) remains higher than raw QT MAE
(17.5~ms), reflecting the sensitivity of Bazett's formula to heart rate
extremes: at high rates, small QT errors are amplified by the division by a
small $\sqrt{\text{RR}}$ term. Differences in correction formula
implementations between reference annotations and our pipeline may also
contribute. Alternative correction formulas (Fridericia, Hodges) would be
expected to reduce this variability and are recommended for future evaluation,
particularly in tachyarrhythmia populations.

\subsection{The Case for Interval-Level Benchmarking}

This paper demonstrates that large-scale interval accuracy evaluation is
feasible, informative, and substantially more clinically relevant than the
fiducial-error benchmarks that currently dominate the delineation literature.
Our stratified analysis---showing dramatically different QT accuracy between
sinus rhythm (MAE 9~ms) and SVT (MAE 75~ms) on the same overall dataset---is
a finding that a single aggregate metric completely obscures. We encourage
future delineation papers to include interval measurement accuracy with
rhythm stratification and Bland-Altman statistics alongside traditional
fiducial-point metrics.

\subsection{Limitations}

\begin{enumerate}
  \item The positive QRS bias ($+$12.6~ms) indicates systematic over-estimation
    and should be addressed via boundary recalibration before clinical
    deployment.
  \item QTc LoA ($-$102 to $+$126~ms) are wide, reflecting Bazett's
    sensitivity to heart rate extremes, T-wave morphology variability, and
    differences between the two automated measurement systems.
  \item QT measurement is unreliable in SVT (MAE 75~ms); the system should
    flag this rhythm class and not report automated QT in this context.
  \item The evaluation compares against GE MUSE automated measurements rather
    than direct cardiologist annotation; prospective clinical validation
    against expert-annotated intervals remains future work.
  \item BUT PDB training data and the MIT-BIH-P test set share a common
    upstream source (MIT-BIH Arrhythmia Database); while splits are
    patient-wise, indirect provenance overlap cannot be fully excluded and
    may marginally inflate external P-wave Dice scores.
  \item PR interval measurement accuracy is not evaluated as the reference
    dataset does not include PR interval annotations; the pipeline implements
    PR measurement as a system capability, but its accuracy remains
    uncharacterised.
  \item For morphologically distorted cases (SVT, bundle branch block, wide-QRS rhythms), the GE MUSE reference measurements are themselves of uncertain reliability. Error figures for these subgroups represent inter-system disagreement rather than true accuracy against a validated ground truth, and should be interpreted accordingly.
\end{enumerate}

\section{Conclusion}

We have presented an automated ECG interval measurement pipeline based on
FFCResNet wave segmentation with full statistical characterisation on 10,646
clinical ECGs measured against GE MUSE commercial reference values. The
primary findings are:

\begin{itemize}
  \item QT MAE 17.5~ms [CI: 16.9--18.2], within manual inter-observer
    variability, with a clinically modest bias of $+$8.5~ms.
  \item QRS MAE 14.8~ms [CI: 14.6--15.0], within clinical tolerance, but
    with a consistent $+$12.6~ms bias requiring recalibration.
  \item Ventricular rate MAE 0.8~bpm, clinically negligible.
  \item Rhythm-stratified analysis shows QT MAE 7--14~ms on sinus rhythms,
    27--36~ms on AF, and 75~ms on SVT---the first such stratification in the
    open delineation literature.
\end{itemize}

The wave segmentation component achieves internal Dice of 95.5--98.2\% and
cross-database Dice of 74.2--85.5\%. The FFC pathway and register tokens
each contribute consistent, modest improvements over a standard ResNet
baseline. Segmentation metrics are presented as system characterisation only,
not in direct numerical comparison with prior methods, which use the
incommensurable fiducial-point timing error metric.

Future work will target: QRS boundary recalibration to eliminate the positive
bias; SVT-specific QT measurement strategies; domain adaptation for improved
cross-database segmentation; multi-lead axis and chamber enlargement criteria;
prospective clinical validation against expert-annotated intervals; and
evaluation of Fridericia and Hodges QTc corrections for tachyarrhythmia
populations.

This work also demonstrates that large-scale interval accuracy evaluation
with rhythm stratification is both feasible and more clinically informative
than the fiducial-error benchmarks that currently dominate the delineation
literature. The dramatically different QT accuracy between sinus rhythm
(MAE 9~ms) and SVT (MAE 75~ms)---obscured by a single aggregate
number---exemplifies why rhythm-stratified interval reporting should become
standard practice.



\begin{thebibliography}{99}

\bibitem{Pan1985}
J.~Pan and W.~J.~Tompkins,
``A real-time QRS detection algorithm,''
\textit{IEEE Trans. Biomed. Eng.}, vol.~BME-32, no.~3, pp.~230--236, 1985.

\bibitem{Kohler2002}
B.-U.~Kohler, C.~Hennig, and R.~Orglmeister,
``The principles of software QRS detection,''
\textit{IEEE Eng. Med. Biol. Mag.}, vol.~21, no.~1, pp.~42--57, 2002.

\bibitem{Martinez2004}
J.~P.~Mart{\'i}nez, R.~Almeida, S.~Olmos, A.~P.~Rocha, and P.~Laguna,
``A wavelet-based ECG delineator: evaluation on standard databases,''
\textit{IEEE Trans. Biomed. Eng.}, vol.~51, no.~4, pp.~570--581, 2004.

\bibitem{Elgendi2014}
M.~Elgendi, B.~Eskofier, S.~Dokos, and D.~Abbott,
``Revisiting QRS detection methodologies for portable, wearable, battery-operated,
and wireless ECG systems,''
\textit{PLoS ONE}, vol.~9, no.~1, p.~e84018, 2014.

\bibitem{Rajpurkar2017}
P.~Rajpurkar \textit{et al.},
``Cardiologist-level arrhythmia detection with convolutional neural networks,''
arXiv:1707.01836, 2017.

\bibitem{Hannun2019}
A.~Y.~Hannun \textit{et al.},
``Cardiologist-level arrhythmia detection and classification in ambulatory
electrocardiograms using a deep neural network,''
\textit{Nat. Med.}, vol.~25, pp.~65--69, 2019.

\bibitem{Ronneberger2015}
O.~Ronneberger, P.~Fischer, and T.~Brox,
``U-Net: Convolutional networks for biomedical image segmentation,''
in \textit{Proc. MICCAI}, 2015, pp.~234--241.

\bibitem{Moskalenko2019}
V.~Moskalenko, N.~Zolotykh, and G.~Osipov,
``Deep learning for ECG segmentation,''
in \textit{Proc. Int. Conf. Neuroinformatics}, 2019, pp.~246--254.

\bibitem{Jimenez2019}
G.~Jimenez-Perez, A.~Alcaine, and O.~Camara,
``U-Net architecture for the automatic detection and delineation of the
electrocardiogram,''
in \textit{Proc. Computing in Cardiology}, 2019, pp.~1--4.

\bibitem{Duraj2022}
K.~Duraj, N.~Piaseczna, P.~Kostka, and E.~Tkacz,
``Semantic segmentation of 12-lead ECG using 1D residual U-Net with
squeeze-excitation blocks,''
\textit{Applied Sciences}, vol.~12, no.~7, p.~3332, 2022.

\bibitem{Joung2024}
C.~Joung, M.~Kim, T.~Paik \textit{et al.},
``Deep learning based ECG segmentation for delineation of diverse arrhythmias,''
\textit{PLoS ONE}, vol.~19, no.~6, p.~e0303178, 2024.

\bibitem{ChenSegNet2022}
Z.~Chen, M.~Wang, M.~Zhang \textit{et al.},
``ECG\_SegNet: An ECG delineation model based on the encoder-decoder structure,''
\textit{Comput. Biol. Med.}, vol.~144, p.~105445, 2022.

\bibitem{SCFNet2025}
``ECG waveform segmentation via dual-stream network with selective context
fusion,''
\textit{Electronics}, vol.~14, no.~19, p.~3925, 2025.

\bibitem{Park2025}
J.~Park, T.~Park, J.-M.~Kwon, and Y.-Y.~Jo,
``Benchmarking ECG delineation using deep neural network-based semantic
segmentation models,''
in \textit{Proc. CHIL}, vol.~287, pp.~63--88, 2025.

\bibitem{Savelieva1998}
I.~Savelieva, G.~Yi, X.~Guo \textit{et al.},
``Agreement and reproducibility of automatic versus manual measurement of QT
interval and QT dispersion,''
\textit{Am. J. Cardiol.}, vol.~81, no.~4, pp.~471--477, 1998.

\bibitem{DeBie2020}
J.~De~Bie \textit{et al.},
``Interchangeability of automated ECG measurements from different vendors,''
\textit{J. Electrocardiology}, vol.~59, pp.~62--67, 2020.

\bibitem{Bland1986}
J.~M.~Bland and D.~G.~Altman,
``Statistical methods for assessing agreement between two methods of clinical
measurement,''
\textit{Lancet}, vol.~327, no.~8476, pp.~307--310, 1986.

\bibitem{Zheng2022}
J.~Zheng, H.~Guo, and H.~Chu,
``A large scale 12-lead electrocardiogram database for arrhythmia study
(version 1.0.0),''
PhysioNet, 2022. DOI: 10.13026/wgex-er52.

\bibitem{Kalyakulina2020}
A.~I.~Kalyakulina, I.~I.~Yusipov, V.~A.~Moskalenko \textit{et al.},
``LUDB: a new open-access validation tool for electrocardiogram delineation
algorithms,''
arXiv:1809.03393, 2020.

\bibitem{Laguna1997}
P.~Laguna, R.~G.~Mark, A.~Goldberg, and G.~B.~Moody,
``A database for evaluation of algorithms for measurement of QT and other
waveform intervals in the ECG,''
in \textit{Computers in Cardiology}, 1997, pp.~673--676.

\bibitem{MarsBUTPDB2021}
L.~Mar\v{s}\'anov\'a, A.~Nemcova, R.~Smisek, L.~Smital, and M.~Vitek,
``Brno University of Technology ECG Signal Database with Annotations of
P Wave (BUT PDB) (version 1.0.0),''
PhysioNet, 2021. DOI: 10.13026/hwvj-5b53.

\bibitem{Avetisyan2025}
A.~Avetisyan, N.~Khachaturov, A.~Asatryan, S.~Tigranyan, and Y.~Markin,
``ISP ECG delineation dataset (version 2),''
Zenodo, 2025. DOI: 10.5281/zenodo.14679837.

\bibitem{Chi2020}
L.~Chi, B.~Jiang, and Y.~Mu,
``Fast Fourier convolution,''
in \textit{Advances in Neural Information Processing Systems (NeurIPS)},
vol.~33, pp.~4479--4490, 2020.

\bibitem{Thygesen2018}
K.~Thygesen \textit{et al.},
``Fourth universal definition of myocardial infarction,''
\textit{Circulation}, vol.~138, pp.~e618--e651, 2018.

\bibitem{Wagner2009}
G.~S.~Wagner,
\textit{Marriott's Practical Electrocardiography}, 11th ed.
Philadelphia: Lippincott Williams \& Wilkins, 2009.

\bibitem{Friesen1990}
G.~M.~Friesen \textit{et al.},
``A comparison of the noise sensitivity of nine QRS detection algorithms,''
\textit{IEEE Trans. Biomed. Eng.}, vol.~37, no.~1, pp.~85--98, 1990.

\bibitem{Chatterjee2020}
S.~Chatterjee, R.~S.~Thakur, R.~N.~Yadav, L.~Gupta, and D.~K.~Raghuvanshi,
``Review of noise removal techniques in ECG signals,''
\textit{IET Signal Process.}, vol.~14, no.~9, pp.~569--590, 2020.

\bibitem{Darcet2023}
T.~Darcet, M.~Oquab, J.~Mairal, and P.~Bourdoukis,
``Vision transformers need registers,''
arXiv:2309.16588, 2023.

\bibitem{Goldberger2018}
A.~L.~Goldberger, Z.~D.~Goldberger, and A.~Shvilkin,
\textit{Goldberger's Clinical Electrocardiography: A Simplified Approach},
9th ed. Philadelphia: Elsevier, 2018.

\bibitem{Rautaharju2009}
P.~M.~Rautaharju \textit{et al.},
``AHA/ACCF/HRS recommendations for the standardization and interpretation
of the electrocardiogram,''
\textit{Circulation}, vol.~119, pp.~e235--e240, 2009.

\bibitem{Postema2008}
P.~G.~Postema and A.~A.~Wilde,
``The measurement of the QT interval,''
\textit{Curr. Cardiol. Rev.}, vol.~4, no.~3, pp.~205--211, 2008.

\bibitem{Kligfield2007}
P.~Kligfield \textit{et al.},
``Recommendations for the standardization and interpretation of the
electrocardiogram,''
\textit{J. Am. Coll. Cardiol.}, vol.~49, pp.~1109--1127, 2007.

\bibitem{Surawicz2009}
B.~Surawicz \textit{et al.},
``AHA/ACCF/HRS recommendations for the standardization and interpretation
of the electrocardiogram,''
\textit{J. Am. Coll. Cardiol.}, vol.~53, pp.~976--981, 2009.

\bibitem{VanAlste1986}
J.~A.~Van~Alste and T.~S.~Schilder,
``Removal of base-line wander and power-line interference from the ECG by
an efficient FIR filter with a reduced number of taps,''
\textit{IEEE Trans. Biomed. Eng.}, vol.~BME-32, no.~12, pp.~1052--1060, 1985.

\bibitem{Thygesen2012}
K.~Thygesen \textit{et al.},
``Third universal definition of myocardial infarction,''
\textit{Circulation}, vol.~126, pp.~2020--2035, 2012.

\bibitem{OGara2013}
P.~T.~O'Gara \textit{et al.},
``2013 ACCF/AHA guideline for the management of ST-elevation myocardial
infarction,''
\textit{Circulation}, vol.~127, pp.~e362--e425, 2013.

\bibitem{Ibanez2018}
B.~Ibanez \textit{et al.},
``2017 ESC Guidelines for the management of acute myocardial infarction
in patients presenting with ST-segment elevation,''
\textit{Eur. Heart J.}, vol.~39, pp.~119--177, 2018.

\bibitem{Nikus2004}
K.~C.~Nikus \textit{et al.},
``ST depression with negative T waves in lead aVR---A marker for high-risk
coronary artery disease,''
\textit{Ann. Noninvasive Electrocardiol.}, vol.~9, no.~3, pp.~207--214, 2004.

\end{thebibliography}
\end{document}